
\magnification=\magstep1
\parskip 0pt
\parindent 15pt
\baselineskip 16pt
\hsize 5.53 truein
\vsize 8.5 truein

\font\titolo = cmbx9 scaled \magstep2
\font\autori = cmsl9 scaled \magstep2
\font\address = cmr9
\font\abstract = cmr8

\hrule height 0pt
\rightline{POLFIS-TH.19/92}
\centerline{\titolo SURFACE REENTRANCE IN THE}
\vskip .05truein
\centerline{\titolo SEMI-INFINITE SPIN-1 ISING MODELS}
\vskip .15truein
\centerline{\autori Carla Buzano and Alessandro Pelizzola}
\vskip .15truein
\centerline{\address Dipartimento di Fisica, Politecnico di Torino,
I-10129 Torino, Italy}

\vfill

\centerline{\bf ABSTRACT}
\vskip .05truein {\address The critical behavior of the semi-infinite
Blume-Capel and Blume-Emery-Griffiths models is investigated in the pair
approximation of the Cluster Variation Method. Equations for bulk and
surface order parameters and n.n. correlation functions are given, from
which analytical expressions for the second order bulk and surface critical
temperatures are derived. The phase diagrams of the Blume-Capel model are
classified, and the existence of a surface first order transition is
discussed. This transition is shown to be, under certain conditions,
slightly reentrant, and the behavior of the surface order parameters and
correlation functions is given for such a case. The extension of our
results to the Blume-Emery-Griffiths model is briefly discussed.}
\vskip .35 truein
\vfill
\eject

\parindent 15pt

{\bf 1. Introduction}
\medskip
The spin-1 Ising models have been originally proposed to describe
the critical behavior of magnetic systems[1], and have subsequently been
applied to He$^3$-He$^4$ mixtures[2] and to multicomponent fluids[3].
The most common and most extensively studied among these models are the
Blume-Capel (BC) model[1] and the Blume-Emery-Griffiths (BEG) model[2], the
former being but a special case, with no biquadratic exchange interaction,
of the latter. \par
These models are not exactly solvable in more than one dimension
(one exception being the BEG model in a particular subspace of the phase
space[4]), but they have
been studied over infinite $d$-dimensional lattices with a lot of different
approximation techniques,
and their phase diagrams are well-known. \par
In the last few years some attention has been devoted to the study of these
models over semi-infinite lattices, with different
couplings at the surface and in the bulk. Benyoussef, Boccara and Saber[5]
have investigated the BC model in the mean field approximation (MFA), giving a
complete classification of the possible types of phase diagrams, while
Benyoussef, Boccara and El Bouziani[6] have carried out the same investigation
on the BEG model
in a real-space renormalization group (RG) framework. A few other
papers have been devoted to the application of the semi-infinite BEG
model to the study of surface superfluidity in He$^3$-He$^4$ mixtures,
both with RG[7,8] and MFA[9], and a particular investigation of the BC model,
in the effective field approximation (EFA), in the case of equal bulk and
surface
exchange interactions is due to Tamura[10]. \par
As in the case of the spin-1/2 Ising model, it is possible to have an
ordered surface even when the bulk is disordered. When such a situation
occurs, it is customary to speak of {\it extraordinary} (i.e. bulk) and
{\it surface} transition, and the temperature of the extraordinary
transition is lower than the surface one. Otherwise, bulk and surface
disorder at the same temperature and with the same critical exponents and the
transition is said to be {\it ordinary}. Finally, in the limiting case
between the two situations above, one has a {\it special} transition, with
bulk and surface disordering at the same temperature but with different
critical exponents. \par
While these results are well-established, the
literature shows a controversy about the order of these transitions.
Some papers, where MFA[5,9] or RG[7,8] are employed, indicate clearly that
the surface transition (and thus also the extraordinary one) can be either
second or first order; on the contrary, there are authors who exclude the
possibility of a first order surface or extraordinary transition, both in a
RG[6] and in an EFA[10] approximation scheme.  \par
In this paper we investigate the semi-infinite spin-1 Ising models in the pair
approximation of the cluster variation method (CVM), which has been
introduced by Kikuchi[11] and has subsequently been reformulated by An[12],
who put it into a much simpler form resorting to the M\"obius inversion. \par
In order to allow for the spatial variation of the local quantities which
this method introduces, we divide the lattice into an infinite set of
layers which are parallel to the free surface:
since this procedure leads to an infinite number of equations, we make another
approximation, which consists in treating the first layer below the surface
as if it was bulk; this way we obtain only two finite sets of
equations: one for the bulk, which is not coupled to the surface one,
and one for the surface, which contains the bulk variables as parameters. \par
The use of the CVM has two main advantages: it allows us to obtain the
correlation functions in a quite straightforward way and to distinguish
easily between second and first order transitions, the former being located
analytically. Furthermore, we overtake the problem of dealing with a great
number of variables and equations by considering a minimal set of
indipendent functions (the order parameters and the correlation functions) of
the elements of the reduced density matrices, and by developing moreover a
procedure which allows us to solve the equations for the correlation functions,
leaving us with only two coupled equations for the order parameters.
These equations, when solved numerically by iteration, give always
thermodinamically stable solutions, i.e. minima of the free energy.
{}From the equations for the order parameters we deduce an
equation for the second order critical temperature, while first order
transitions are determined numerically. \par
The analysis, in the whole phase space of the BC model, shows two new types
of phase diagrams and a reentrant first order surface transition. For the
latter, we give qualitative conditions about the region of the phase space
where it occurs and an example of behavior of the order parameters and of
the correlation functions. \par
The paper is organized as follows: in Sec. 2 we determine the CVM free
energy, the equations for the order parameters and the second order
critical temperatures; in Sec. 3 we describe our main results: the
classification of the phase diagrams for the BC model and the
analysis of the reentrant phenomenon in the first order surface transition.
In Sec. 4 we discuss the
extension of our results to the BEG model and some conclusions are drawn in
Sec. 5. \par
\bigskip
{\bf 2. The model and the approximation.}
\medskip
The semi-infinite BEG model has hamiltonian
$$\eqalign{
\beta H_{BEG} = & \quad -J_S \sum_{\langle i j \rangle} S_i S_j
-J_B \sum_{\langle k l \rangle} S_k S_l
+\Delta_S \sum_i S_i^2 + \Delta_B \sum_k S_k^2 \cr
& - K_S \sum_{\langle i j \rangle} S_i^2 S_j^2
- K_B \sum_{\langle k l \rangle} S_k^2 S_l^2 \cr}, \eqno(1)$$
where $i,j,k,l$ are site labels,
$S_i$ is the $z$-component of a spin-1 operator at site $i$,
${\displaystyle \sum_{\langle i j \rangle}}$ denotes a sum over all
nearest neighbors (n.n.) with both sites lying on the surface,
${\displaystyle \sum_{\langle k l \rangle}}$ denotes a sum over the remaining
n.n., and $\beta = (k_B T)^{-1}$ (with $k_B$ Boltzmann constant and $T$
absolute temperature).
$J_S$ and $J_B$ (both positive, since we study the ferromagnetic case) are
reduced exchange interactions, while $\Delta_S$ and $\Delta_B$ are reduced
crystal fields and $K_S$ and
$K_B$ are reduced biquadratic exchange interactions,
respectively at the surface and in the bulk. \par
As we mentioned in the introduction, we divide our lattice into layers,
parallel to the
surface, labeled by an integer $n$, $n = 1$ being the surface layer. For
the sake of simplicity in the following we will consider a simple cubic
lattice with a (100) free surface. \par
The CVM is based on an approximate expression of the entropy of the model as a
sum of contributions by all the elements of a set ${\cal M}$ of maximal
clusters and all their subclusters. In the pair approximation, one chooses
as ${\cal M}$ the set of all n.n. pairs. \par
Thus our entropy will be the sum of contributions of three different kinds:
one due to one-site clusters on layer $n$, denoted by $\sigma_s^{(n)}$;
one from two-site clusters with both sites in layer $n$, denoted by
$\sigma_p^{(n)}$ and
one from two-site clusters with a site in layer $n$ and another in layer
$n+1$, denoted by $\sigma_{p^\prime}^{(n)}$.
All these contributions will be weighted appropriately yielding, for the
total entropy $\sigma$
$$\sigma/N_S = \sum_\gamma a_\gamma (N_\gamma/N_S) \sigma_\gamma,
\eqno(2)$$
where $N_S$ is the number of sites in a layer and $\gamma$ ranges over all
the clusters above. The weights $a_\gamma$ can be determined by means of
An's equations[12] and, as well as the multiplicities $N_\gamma/N_S$, are
lattice dependent. In our case we obtain
$$a_s^{(1)} = -4, \qquad a_s^{(n)} = -5, \quad n \ge 2, \qquad
N_s^{(n)}/N_S = 1, \quad n \ge 1$$
$$a_p^{(n)} = a_{p^\prime}^{(n)} = 1, \qquad N_p^{(n)}/N_S = 2, \qquad
N_{p^\prime}^{(n)}/N_S = 1, \qquad n \ge 1.\eqno(3)$$
Finally, $\sigma_\gamma$ is the entropy associated to the cluster $\gamma$
and is given by
$$\sigma_\gamma = - k_B {\rm Tr}(\rho_\gamma \ln \rho_\gamma), \eqno(4)$$
where $\rho_\gamma$ is the reduced density matrix for cluster $\gamma$,
which has to be determined by minimization of the free energy. \par
Since the model hamiltonian is diagonal the reduced density matrices turn
out to be diagonal as well. Furthermore, they must obey the following
constraints:
$${\rm Tr}\rho_\gamma = 1, \qquad
\sum_\epsilon {\rho_p^{(n)}}_{\delta\epsilon} =
\sum_\epsilon {\rho_p^{(n)}}_{\epsilon\delta} =
{\rho_s^{(n)}}_\delta$$
$$\sum_\epsilon {\rho_{p^\prime}^{(n)}}_{\delta\epsilon} =
{\rho_s^{(n)}}_\delta, \qquad
\sum_\epsilon {\rho_{p^\prime}^{(n)}}_{\epsilon\delta} =
{\rho_s^{(n+1)}}_\delta,\eqno(5)$$
where ${\rho_p^{(n)}}_{\delta\epsilon} = \langle \delta \epsilon \vert
\rho_p^{(n)} \vert \delta \epsilon \rangle$ and so on, with
$\delta$ and $\epsilon$ taking values $+,0,-$, and
it is assumed for $\rho_{p^\prime}^{(n)}$ that the first index refer
to the site in layer $n$ and the second one to the site in layer $n + 1$. \par
These constraints allow us to re-express the density matrices as functions
of a reduced (with respect to the set of the elements of the matrices
theirselves) set of order parameters and correlation functions; this is the
first step in lowering the number of equations one must deal with when
minimizing the free energy. Upon defining the order parameters ($\langle
\quad \rangle$ denotes now thermal average and $S_i^{(n)}$ stands for a spin in
the $n$th layer)
$$y_1^{(n)} = \langle S_i^{(n)} \rangle \quad {\rm and} \quad
y_2^{(n)} = \langle {S_i^{(n)}}^2 \rangle, \eqno(6)$$
and the n.n. two-points correlation functions
$$\eqalign{& y_3^{(n)} = \langle S_i^{(n)} S_j^{(n)} \rangle, \qquad
y_3^{\prime(n)} = \langle S_i^{(n)} S_j^{(n+1)} \rangle; \cr
& y_4^{(n)} = \langle S_i^{(n)} {S_j^{(n)}}^2 \rangle, \qquad
y_4^{\prime(n)} = \langle S_i^{(n)} {S_j^{(n+1)}}^2 \rangle, \qquad
y_4^{\prime\prime(n)} = \langle {S_i^{(n)}}^2 S_j^{(n+1)} \rangle; \cr
& y_5^{(n)} = \langle {S_i^{(n)}}^2 {S_j^{(n)}}^2 \rangle, \qquad
y_5^{\prime(n)} = \langle {S_i^{(n)}}^2 {S_j^{(n+1)}}^2 \rangle. \cr}
\eqno(7)$$
one obtains the following expressions for the elements of the density matrices:
$$ {\rho_s^{(n)}}_\pm = {y_2^{(n)} \pm y_1^{(n)} \over 2}, \qquad
{\rho_s^{(n)}}_0 = 1 - y_2^{(n)}; $$

$${\rho_p^{(n)}}_{\delta \epsilon} =
{y_5^{(n)} + \delta \epsilon y_3^{(n)} + (\delta + \epsilon) y_4^{(n)}
\over 4}, \qquad
{\rho_p^{(n)}}_{00} =
1 + y_5^{(n)} - 2 y_2^{(n)}, $$
$${\rho_p^{(n)}}_{\pm 0} =
{\rho_p^{(n)}}_{0 \pm} =
{y_2^{(n)} - y_5^{(n)} \pm y_1^{(n)} \mp y_4^{(n)} \over 2};$$

$$ {\rho_{p^\prime}^{(n)}}_{\delta \epsilon} =
{y_5^{\prime(n)} + \delta\epsilon y_3^{\prime(n)} + \delta y_4^{\prime(n)}
+ \epsilon y_4^{\prime\prime(n)} \over 4}, $$
$${\rho_{p^\prime}^{(n)}}_{\pm 0} =
{y_2^{(n)} - y_5^{\prime(n)} \pm y_1^{(n)} \mp y_4^{\prime(n)} \over 2},$$
$${\rho_{p^\prime}^{(n)}}_{0 \pm} =
{y_2^{(n+1)} - y_5^{\prime(n)} \pm y_1^{(n+1)} \mp
y_4^{\prime\prime(n)} \over 2},$$
$${\rho_{p^\prime}^{(n)}}_{0 0} =
1 + y_5^{\prime(n)} - y_2^{(n)} - y_2^{(n+1)},\eqno(8)$$
where now $\delta$ and $\epsilon$ take values $+,-$. \par

We are now able to write down our approximate expression for the reduced free
energy density $f = \beta (U - T\sigma)/N_S$;
we obtain
$$\eqalign{f = \ & \Delta_S \ y_2^{(1)}
- 2 K_S \ y_5^{(1)}
- 2 J_S \ y_3^{(1)} \cr
& + \sum_{n=2}^\infty \left[ \Delta_B \ y_2^{(n)}
- 2 J_B \ y_3^{(n)}
- 2 K_B \ y_5^{(n)} \right] \cr
& - \sum_{n=1}^\infty \left[ J_B \ y_3^{\prime(n)} + K_B \ y_5^{\prime(n)}
\right] \cr
& + \sum_{n=1}^\infty \left[ 2
{\rm Tr} \left( \rho_p^{(n)} \ln \rho_p^{(n)} \right) +
{\rm Tr} \left( \rho_{p^\prime}^{(n)} \ln \rho_{p^\prime}^{(n)} \right)
\right] \cr
& - 4 {\rm Tr} \left( \rho_s^{(1)} \ln \rho_s^{(1)} \right)
- 5 \sum_{n=2}^\infty
{\rm Tr} \left( \rho_s^{(n)} \ln \rho_s^{(n)} \right). \cr} \eqno(9)$$
Requiring that $f$ is a minimum with respect to all the $y$'s yields an
infinite set of coupled equations, which we report making
use of the notation $\lambda_s^{(n)} = \ln \rho_s^{(n)}$,
$\lambda_p^{(n)} = \ln \rho_p^{(n)}$, $\lambda_{p^\prime}^{(n)} =
\ln \rho_{p^\prime}^{(n)}$, and indices have the same meaning as in (8):
\bigskip
\leftline{$ 4({\lambda_p^{(n)}}_{+  0} - {\lambda_p^{(n)}}_{- 0}) + $}
\smallskip
\leftline{$
({\lambda_{p^\prime}^{(n)}}_{+ 0} - {\lambda_{p^\prime}^{(n)}}_{- 0} +
{\lambda_{p^\prime}^{(n-1)}}_{0 +} - {\lambda_{p^\prime}^{(n-1)}}_{0 -}) +
a_s^{(n)} ({\lambda_s^{(n)}}_{+} - {\lambda_s^{(n)}}_{-}) = 0$}
\bigskip
\leftline{$2D_n + 4({\lambda_p^{(n)}}_{+ 0} + {\lambda_p^{(n)}}_{- 0} -
2{\lambda_p^{(n)}}_{0 0}) +$}
\smallskip
\leftline{$({\lambda_{p^\prime}^{(n)}}_{+ 0} +
{\lambda_{p^\prime}^{(n)}}_{- 0} - 2{\lambda_{p^\prime}^{(n)}}_{0 0} +
{\lambda_{p^\prime}^{(n-1)}}_{0 +} + {\lambda_{p^\prime}^{(n-1)}}_{0 -} -
2{\lambda_{p^\prime}^{(n-1)}}_{0 0}) + $}
\smallskip
\leftline{$a_s^{(n)}
({\lambda_s^{(n)}}_{+} + {\lambda_s^{(n)}}_{-} - 2{\lambda_s^{(n)}}_{0}) = 0$}
\bigskip
\leftline{$-4J_n + ({\lambda_p^{(n)}}_{+ +} + {\lambda_p^{(n)}}_{- -} -
2{\lambda_p^{(n)}}_{+ -}) = 0$}
\bigskip
\leftline{$-4J_B + ({\lambda_{p^\prime}^{(n)}}_{+ +} +
{\lambda_{p^\prime}^{(n)}}_{- -} - {\lambda_{p^\prime}^{(n)}}_{+ -} -
{\lambda_{p^\prime}^{(n)}}_{- +}) = 0$}
\bigskip
\leftline{${\lambda_p^{(n)}}_{+ +} - {\lambda_p^{(n)}}_{- -} -
2({\lambda_p^{(n)}}_{+ 0} - {\lambda_p^{(n)}}_{- 0}) = 0$ \hfill (10)}
\bigskip
\leftline{$({\lambda_{p^\prime}^{(n)}}_{+ +} +
{\lambda_{p^\prime}^{(n)}}_{+ -} - {\lambda_{p^\prime}^{(n)}}_{- +} -
{\lambda_{p^\prime}^{(n)}}_{- -}) -
2({\lambda_{p^\prime}^{(n)}}_{+ 0} - {\lambda_{p^\prime}^{(n)}}_{- 0}) = 0$}
\bigskip
\leftline{$({\lambda_{p^\prime}^{(n)}}_{+ +} -
{\lambda_{p^\prime}^{(n)}}_{+ -} + {\lambda_{p^\prime}^{(n)}}_{- +} -
{\lambda_{p^\prime}^{(n)}}_{- -}) -
2({\lambda_{p^\prime}^{(n)}}_{0 +} - {\lambda_{p^\prime}^{(n)}}_{0 -}) = 0$}
\bigskip
\leftline{$-4K_n + ({\lambda_p^{(n)}}_{+ +} + {\lambda_p^{(n)}}_{- -} +
2{\lambda_p^{(n)}}_{+ -}) + 4{\lambda_p^{(n)}}_{0 0} - $}
\smallskip
\leftline{$ 4({\lambda_p^{(n)}}_{+ 0} + {\lambda_p^{(n)}}_{- 0}) = 0$}
\bigskip
\leftline{$- 4K_B + ({\lambda_{p^\prime}^{(n)}}_{+ +} +
{\lambda_{p^\prime}^{(n)}}_{+ -} + {\lambda_{p^\prime}^{(n)}}_{- +} +
{\lambda_{p^\prime}^{(n)}}_{- -} + {\lambda_{p^\prime}^{(n)}}_{0 0} -$}
\smallskip
\leftline{$2({\lambda_{p^\prime}^{(n)}}_{+ 0} +
{\lambda_{p^\prime}^{(n)}}_{- 0} + {\lambda_{p^\prime}^{(n)}}_{0 +} +
{\lambda_{p^\prime}^{(n)}}_{0 -}) = 0. $}
\bigskip
Here $n$ ranges from 1 to $\infty$, $J_1 = J_S, J_n = J_B$ for $n \ge 2$
and similarly for $D_n$ and $K_n$, and $\lambda_{p^\prime}^{(0)} = 0$. \par
Letting $n$ go to $\infty$ in the equations above one obtains the bulk
equations, which are only five, since in the bulk correlations like $y_3$
and $y_3^\prime$ coincide. We have already solved the bulk equations in
Ref. 13 and
here we report the two coupled equations for the bulk order parameters
$$\eqalign{
& y_1^{(B)} = \left[ (V_+ - V_-) + \eta(V_+^2 - V_-^2) \right]/W \cr
& y_2^{(B)} = \left[ (V_+ + V_-) + \eta(V_+^2 + V_-^2) + 2\gamma V_+V_-
\right]/W
\cr} \eqno(11)$$
and the expressions for the correlation functions given as functions of the
order parameters
$$\eqalign{
& y_3^{(B)} = \left[ \eta(V_+^2 + V_-^2) - 2\gamma V_+V_- \right]/W \cr
& y_4^{(B)} = \left[ \eta(V_+^2 - V_-^2) \right]/W \cr
& y_5^{(B)} = \left[ \eta(V_+^2 + V_-^2) + 2\gamma V_+V_- \right]/W \cr}
\quad , \eqno(12)$$
where $\ \eta = \exp(J_B + K_B), \ \gamma = \exp(-J_B + K_B),$
$$V_\pm = e^{-\Delta_B/6}\left[{y_2^{(B)} \pm y_1^{(B)} \over 2(1 - y_2^{(B)})}
\right]^{5 \over 6}, \eqno(13)$$
$W = \eta(V_+^2 + V_-^2) + 2\gamma V_+V_- + 2(V_+ + V_-) + 1$ and
$y_i^{(B)} = {\displaystyle \lim_{n \to \infty}} y_i^{(n)}$. \par
{}From equations (11) one can easily derive the equation for the second
order critical temperature, which reads[13]
$$e^{\Delta_B} = 2 \zeta (\gamma_0 - 1)
\left[ {\zeta \gamma_0 \over \zeta (\gamma_0 - 1) + 1} \right]^5,\eqno(14)$$
where
$\zeta = e^{K_B}\cosh J_B$ and $\gamma_0 = 5 \tanh J_B.$ \par
Turning to the whole set of equations (10) let us observe that, since is
not possible to decouple different layers,
some sort of approximation is in order; one
can, for example, choose a maximum number $\bar n$ of layers and then solve
numerically the equations for $n = 1, 2, \ldots \bar n$ using the bulk
solutions as boundary conditions, i.e. as if they were the solutions for
layer $\bar n + 1$. We choose the crudest approximation, $\bar n = 1$,
which allows us to obtain analytical results for the second order
transition. The same assumption has been made in Refs. 5-8. \par
The procedure we follow to determine the equations for the surface ($n =
1$) order parameters (which is analogue to that used in [13] for the bulk ones)
can easily be shown to be equivalent to the natural iteration method (NIM)
by Kikuchi[14], in the sense that the resulting equations, when solved by
numerical iteration, give always thermodinamically stable solutions.
Moreover, our procedure has the advantage of dealing with a considerably
lower number of equations (only two, in the present case). \par
{}From now on we will suppress the index $^{(1)}$, which has become redundant;
furthermore, instead of $^{(2)}$ we will write $^{(B)}$, because layer 2
now plays the role of the bulk. \par
In order to determine the surface equations we express
$\rho_{p^\prime} \equiv \rho_{p^\prime}^{(1)}$ and $\rho_p \equiv
\rho_p^{(1)}$ as functions of $\rho_s \equiv \rho_s^{(1)}$ and of the bulk
quantities; to this aim we consider a set of equations formed by the four
equations in the set (10) for $n = 1$ which contain only elements of
$\lambda_{p^\prime}$, the first two equations of that set for $n = 2$, and
the three equations given by the first of the constraints (5) for
$\rho_{p^\prime}$. This way we obtain a system which (recalling that
$\lambda_{p^\prime} = \ln \rho_{p^\prime}$) can be solved for the elements
of $\rho_{p^\prime}$, obtaining
$${\rho_{p^\prime}}_{\delta \epsilon} = \exp \left[ \delta \epsilon J_B +
(\delta \epsilon)^2 K_B \right] {c_\epsilon \over d_\delta}
{\rho_s}_\delta, \qquad \delta,\epsilon = +,0,-, \eqno(15)$$
where
$$c_\pm = e^{-\Delta_B/6} \left( {{\rho_s^{(B)}}_\pm \over {\rho_s^{(B)}}_0}
\right)^{5/6}, \qquad c_0 = 1 \eqno(16)$$
and
$$d_\delta = \sum_\epsilon \exp \left[ \delta \epsilon J_B +
(\delta \epsilon)^2 K_B \right] c_\epsilon \qquad \delta = +,0,-. \eqno(17)$$
The remaining five equations for $n = 1$, together with the normalization
condition ${\rm Tr}(\rho_p) = 1$ can now be used to obtain the
elements of $\rho_p$; one finds
$${\rho_p}_{\delta \epsilon} = \exp \left[ \delta \epsilon J_S +
(\delta \epsilon)^2 K_S \right] \gamma_\delta \gamma_\epsilon
G^{-1}, \qquad \delta,\epsilon = +,0,-, \eqno(18)$$
where
$$\gamma_\pm = e^{-\Delta_S/4} \left( {{\rho_s}_\pm \over {\rho_s}_0}
\right)^{3/4}
\left( {d_\pm \over d_0} \right)^{1/4}, \qquad \gamma_0 = 1
\eqno(19)$$
and
$$G = \sum_{\delta \epsilon} \exp \left[ \delta \epsilon J_S +
(\delta \epsilon)^2 K_S \right] \gamma_\delta \gamma_\epsilon. \eqno(20)$$
Since the surface order parameters are related to $\rho_p$ by the equations
$$\eqalign{
& y_1 = {\rho_p}_{+ +} - {\rho_p}_{- -} + {\rho_p}_{+ 0} - {\rho_p}_{- 0} \cr
& y_2 = {\rho_p}_{+ +} + {\rho_p}_{- -} + {\rho_p}_{+ 0} + {\rho_p}_{- 0}
+ 2{\rho_p}_{+ -} \cr} \eqno(21)$$
they must satisfy the following equations
$$\eqalign{
& y_1 = \left[ e^{J_S + K_S} (\gamma_+^2 - \gamma_-^2) + (\gamma_+ - \gamma_-)
\right] /G \cr
& y_2 = \left[ e^{J_S + K_S} (\gamma_+^2 + \gamma_-^2) +
2e^{-J_S + K_S} \gamma_+ \gamma_- + (\gamma_+ + \gamma_-)
\right] /G \cr}. \eqno(22)$$ \par
Once one has solved (22), using (8), (15) and (18) it is immediate to obtain
explicit expressions for the n.n. two-points correlation functions. \par
It is easy to check that, when $y_1 = 0$, then $y_1^{(B)} = 0$ too, that
is, if the surface is paramagnetic, the bulk must also be paramagnetic. On
the contrary, if the bulk is paramagnetic ($y_1^{(B)} = 0$), the equation
for $y_1$ has always the paramagnetic solution $y_1 = 0$, but can also have
a ferromagnetic solution $y_1 \ne 0$. It means that in the paramagnetic
region of the bulk phase diagram one can have a surface transition line.
The first order part of this line must be evaluated numerically, by
comparison of the free energies of the two phases, while the second order
one can be determined analytically by expanding (22) in powers of $y_1$
(the order parameter $y_2$ must be expanded as well, and one must remember
that $y_1^{(B)} = 0$, because we are in the bulk paramagnetic region of the
phase diagram)
around the solution $y_1 = 0$, and then requiring that
the resulting equations are satisfied to the 3rd
order in $y_1$ (going to the 5th order would give a set of equations for
the tricritical point). One finds, for the
surface second order transition
$$2^{-8} e^{K_S} {c \over c_0} x \left[ 3
{3x + \sqrt{x^2 + 8} \over e^{-K_S} + x} \right]^3 = e^{\Delta_S}, \eqno(23)$$
where $x = e^{J_S} - 2e^{-J_S} > 0$, $c_0 = 1 + 2V$ and
$c = 1 + 2V e^{K_B} \cosh J_B$, with
$$V = e^{-\Delta_B/6}
\left[ {y_2^{(B)} \over 2 (1 - y_2^{(B)})} \right]^{5 \over 6}. \eqno(24)$$
\bigskip
{\bf 3. Phase diagram and reentrant phenomenon}
\medskip
In the present section we turn to the Blume-Capel model: thus we set $K_B = K_S
= 0$. \par
Following [5]  we define the ratios $R = J_B/J_S$ and $D =
\Delta_B/\Delta_S$ and classify the possible phase diagrams at fixed $R,D$
in the plane $(d,\tau)$, where $d = \Delta_B/6J_B$ and $\tau = 1/6J_B$. As
in [5] we obtain four main types of phase diagrams, which we report
in Figs. 1-4; solid and dashed lines represent respectively
second and first order transitions, and the thinner lines, when present, refer
to the surface. B and S stand for bulk and surface respectively, while P
and F stand for para- and ferromagnetic. \par
The type-A diagram, reported in Fig. 1, is characterized by the presence of
only ordinary transitions; the type-B diagram (Fig. 2) shows an ordinary
transition at high temperatures, while in the low temperature region
one has extraordinary and surface transitions: the three transition lines
meet at the special point $X$;
in the type-C diagram (Fig. 3) the situation is reversed, and
finally, in the type-D diagram (Fig. 4) one has no ordinary transition but
two non-intersecting, extraordinary and surface, transition lines. Type-B
and type-C diagrams could be classified with even more detail, because in both
cases the surface tricritical point can be either present or absent, but we
do not enter into these point. \par
It is possible to distinguish between these four types by means of analytic
conditions. First of all one must determine the bulk and surface critical
temperatures at $d = -\infty$, respectively denoted by $\tau_B^{(\infty)}$
and $\tau_S^{(\infty)}$ (the latter is determined under the assumption of
paramagnetic bulk), and the bulk and surface critical values of $d$ at
$\tau = 0$, respectively denoted by $d_B^{(0)}$ and $d_S^{(0)}$, and then
compare these values, recalling that, if $\tau_B^{(\infty)} >
\tau_S^{(\infty)}$ or $d_B^{(0)} > d_S^{(0)}$, the corresponding transition
must be necessarily an ordinary one, because in the bulk ferromagnetic
region of the phase diagram the surface is always ferromagnetic and surface
transitions are forbidden. \par
By means of (14) and (23) one obtains $\tau_B^{(\infty)} = (3\ln{3 \over
2})^{-1} \approx 0.822$ and $\tau_S^{(\infty)} = (3R\ln 2)^{-1} \approx
0.481R^{-1}$ respectively; the condition $\tau_B^{(\infty)} =
\tau_S^{(\infty)}$ gives the critical value
$$R_c = {\ln 3 \over \ln 2} - 1 \approx 0.585, \eqno(25)$$
above which the $d = -\infty$ transition is ordinary (the corresponding MFA
value is $R_c = 2/3$).
Furthermore, simple analytic considerations on the ground state yield
$d_B^{(0)} = 1/2$ and $d_S^{(0)} = {1 \over 3}DR^{-1}$, from which the critical
value
$$D_c \equiv D_c(R) = {3 \over 2}R \eqno(26)$$
(equal to the corresponding MFA value),
below which the zero temperature transition is ordinary, can be derived. Thus,
the four main phase diagrams are characterized by the following conditions:
\item{} Type-A diagram: $R > R_c$ and $D < D_c$,
\item{} Type-B diagram: $R > R_c$ and $D > D_c$,
\item{} Type-C diagram: $R < R_c$ and $D < D_c$,
\item{} Type-D diagram: $R < R_c$ and $D > D_c$. \parindent 0pt \par
\parindent 15pt
Notice that type-B and type-C have not been found in the RG-based
classification of Ref. 6, although a different renormalization group
scheme[7,8] gives a phase diagram which is similar to our type B. \par
To this fundamental classification we must add two new cases, which appear
in the CVM treatment when $D$ is just below $D_c$ and are illustrated
in Figs. 5-6 for $R$
above and just below $R_c$, respectively. The new feature of these diagrams
is the presence of two or more special (i.e. multicritical points where
surface and bulk transition lines meet) points. \par
In Figs. 2,4 and 5 it is possible to observe a reentrant phenomenon in the
ferromagnetic-paramagnetic (order-disorder) surface transition.
In fact, for fixed $R$, there is a range of values of $D$ for which one has
such a phenomenon. The lower and upper limits depend on $R$, and the former is,
of course, just below $D_c$. In Figs. 7-10 we report the behavior of the
order parameters and of the correlation functions for a set of values of
the parameters corresponding to the dotted line in Fig. 4. As can be
expected, the surface-bulk n.n. correlation functions are very small (even
if non-zero) in the temperature range where the bulk is paramagnetic and
the surface ferromagnetic. \par
Finally, it is noteworthy that the surface reentrant phenomenon is not revealed
by
a MFA analysis: in Fig. 11 we compare the MFA phase diagram corresponding to
Fig. 4 to that obtained in our approximation. \par
\bigskip
{\bf 4. The surface reentrance in the BEG model}
\medskip
Let us now briefly consider the BEG model in the case $\xi > -1$, being
$\xi = K_B/6J_B$ (the case $\xi < -1$, where a staggered quadrupolar phase
can occur, would require the introduction of two sublattices, with different
order parameters to allow for the symmetry breaking between them).
The results of the previous section
can be easily extended to this case; the only change
that is needed in the classification is that one finds, for the critical
value of the ratio of crystal fields,
$$D_c = {3 \over 2}R {1 + \xi \over 1 + \xi R \Gamma^{-1}}, \eqno(27)$$
where $\Gamma = K_B/K_S$. Furthermore, since for the BEG
model the bulk phase diagram can exhibit reentrance (see [13] and
references therein) both for $\xi < 0$ and for $\xi > 3$, it is interesting
to know what happens in these cases for the semi-infinite model. \par
For $\xi < 0$ one can easily find cases of reentrant ordinary
transitions, where the bulk and the surface are simultaneously interested by
the reentrant phenomenon, as in Fig. 12. \par
The most interesting
situations occur for $\xi > 3$. When $D$ is slightly less than $D_c$ the
surface reentrant transition already seen for the BC model, together with
the ordinary reentrant transition, give rise to a double reentrant
phenomenon, as shown in Fig. 13. Another double reentrant phenomenon can be
obtained for $D$ slightly greater than $D_c$: in this case (Fig. 14) one
has two successive reentrances on the surface transition line, the highest
temperature transition being now second order. \par
\bigskip
{\bf 5. Final remarks}
\medskip
We have investigated the semi-infinite spin-1 Ising models, devoting
particular attention to the BC model, in the framework of the pair
approximation of the CVM. We have classified the possible phase diagrams at
fixed $R,D$, finding two new types of diagrams and showing that, at least in
the present approximation, the first order surface transition do occur and
exhibits
reentrant phenomenon; finally, we have given qualitative conditions for
this phenomenon and have discussed the extension of our results to the BEG
model. \par
The surface reentrant phenomenon we have found in the BC model is not
revealed by a MFA analysis. It would be interesting to check our results by
other high-precision methods. \par
Our analysis has led to a classification scheme with four fundamental types
of phase diagram, depending on the values of $R$ and $D$. Two of them,
namely type-B and type-C, have not been found in the approximations of
Refs. 6 and 10, while Refs. 7 and 8 present a phase diagram like our type-B.
This point is worth of a further investigation, in order to
establish whether the first order surface transition
exists or it is only an effect due to the approximation method adopted. We
conjecture that this transition is a real feature of the model, and give
the following argument in support. \par
Let us consider the BC model again: at $T = 0$, and for $d > 1/2$, the bulk
ground state is paramagnetic, i.e. $S_k = 0$ for all bulk sites $k$. Then
all the interaction terms between bulk and surface vanish and the surface
behaves like a BC model on an infinite square lattice, thus exhibiting a
first order transition (if $D > D_c$) at $d = d_S^{(0)}$
(indeed, this zero temperature transition is found also in [5]). Similarly, at
a very low but finite temperature, only a small fraction of the bulk spins
will be different from zero and the surface will be only weakly coupled to
the bulk, this resulting in a slight change of the critical value of $d$,
giving a smooth surface first order transition line in the $(d,\tau)$ plane
starting from the point $(d_S^{(0)}, 0)$. Work is in progress to confirm the
conjecture and the results we have obtained. \par
\vfill \eject
{\bf References}
\item{[1]} M. Blume, Phys. Rev. 141 (1966) 517; H.W. Capel, Physica 32 (1966)
966.
\item{[2]} M. Blume, V.J. Emery and R.B. Griffiths, Phys. Rev. A4 (1971) 1071.
\item{[3]} D. Mukamel and M. Blume, Phys. Rev. A10 (1974) 610.
\item{[4]} T. Horiguchi, Phys. Lett. 113A (1986) 425.
\item{[5]} A. Benyoussef, N. Boccara and M. Saber, J. Phys. C19 (1986) 1983.
\item{[6]} A. Benyoussef, N. Boccara and M. El Bouziani, Phys. Rev. B34
(1986) 7775.
\item{[7]} L. Peliti and S. Leibler, J. Physique Lett. 45 (1984) L-596.
\item{[8]} A. Crisanti and L. Peliti, J. Phys. A18 (1985) L543.
\item{[9]} X.P. Jiang and M.R. Giri, J. Phys. C21 (1988) 995.
\item{[10]} I. Tamura, J. Phys. Soc. Jpn. 51 (1982) 3607.
\item{[11]} R. Kikuchi, Phys. Rev. 81 (1951) 988.
\item{[12]} G. An, J. Stat. Phys. 52 (1988) 727.
\item{[13]} C. Buzano and A. Pelizzola, Physica A (in press).
\item{[14]} R. Kikuchi, J. Chem. Phys. 60 (1974) 1071.
\vfill \eject

{\bf Figure captions} \par
\parindent .4 truein
\item{}
\itemitem{\hbox to .67 truein{Fig. 1 : \hfill}}Phase diagram of the
semi-infinite BC model for $R > R_c$
and $D < D_c$. Solid and dashed lines denote second and first order
transitions, respectively; $X$ is the special point; B = bulk, S = surface,
P = paramagnetic, F = ferromagnetic.
\itemitem{\hbox to .67 truein{Fig. 2 : \hfill}}Phase diagram for $R = 1$
and $D = 1.8$. Symbols as in Fig. 1.
\itemitem{\hbox to .67 truein{Fig. 3 : \hfill}}Phase diagram for $R = 0.4$
and $D = 0.5$. Symbols as in Fig. 1.
\itemitem{\hbox to .67 truein{Fig. 4 : \hfill}}Phase diagram for $R = 0.4$
and $D = 0.8$. Symbols as in Fig. 1.
\itemitem{\hbox to .67 truein{Fig. 5 : \hfill}}Phase diagram for $R = 1$
and $D = 1.49$. Symbols as in Fig. 1.
\itemitem{\hbox to .67 truein{Fig. 6 : \hfill}}Phase diagram for $R = 0.56$
and $D = 0.8$. Symbols as in Fig. 1.
\itemitem{\hbox to .67 truein{Fig. 7 : \hfill}}Behavior of the order
parameters $y_1$ (solid line) and $y_2$ (dashed line) for the case
corresponding to the dotted line in Fig. 4.
\itemitem{\hbox to .67 truein{Fig. 8 : \hfill}}Behavior of $y_3$ (solid
line) and $y_3^\prime$ (dashed line) in the case of Fig. 7.
\itemitem{\hbox to .67 truein{Fig. 9 : \hfill}}Behavior of $y_4$ (solid
line), $y_4^\prime$ (dashed line) and $y_4^{\prime\prime}$ (dot-dashed
line) in the case of Fig. 7.
\itemitem{\hbox to .67 truein{Fig. 10 : \hfill}}Behavior of $y_5$ (solid
line) and $y_5^\prime$ (dashed line) in the case of Fig. 7.
\itemitem{\hbox to .67 truein{Fig. 11 : \hfill}}Comparison between the
phase diagram of Fig. 4 (denoted by CVM) and the corresponding one obtained
in MFA (denoted by MFA).
\itemitem{\hbox to .67 truein{Fig. 12 : \hfill}}Phase diagram of
the BEG model for $\xi = -0.5$, $R = 0.5$, $D = 0.4$ and $\Gamma = 1$.
\itemitem{\hbox to .67 truein{Fig. 13 : \hfill}}Phase diagram of
the BEG model for $\xi = 3.6$, $R = 0.5$, $D = 1.22$ and $\Gamma = 1$.
\itemitem{\hbox to .67 truein{Fig. 14 : \hfill}}Phase diagram of
the BEG model for $\xi = 3.6$, $R = 0.5$, $D = 1.26$ and $\Gamma = 1$.
\vfill \eject
\end